\documentclass[conference]{IEEEtran}
\IEEEoverridecommandlockouts
\usepackage{cite}
\usepackage{amsmath,amssymb,amsfonts}
\usepackage{algorithmic}
\usepackage{graphicx}
\usepackage{textcomp}
\usepackage{xcolor}
\usepackage{subcaption}
\usepackage{url}
\usepackage{times}
\usepackage{enumitem}

\def\BibTeX{{\rm B\kern-.05em{\sc i\kern-.025em b}\kern-.08em
    T\kern-.1667em\lower.7ex\hbox{E}\kern-.125emX}}
    
\usepackage{pgfplots}
\pgfplotsset{compat=1.3}
\usepackage{tikz}
\usepackage[lined,linesnumbered]{algorithm2e}

\definecolor{r1}{RGB}{230,185,181}
\definecolor{r2}{RGB}{217,150,144}
\definecolor{r3}{RGB}{192,80,70}
\definecolor{r4}{RGB}{147,58,50}
\definecolor{r5}{RGB}{98,39,33}

\definecolor{b1}{RGB}{183,222,232}
\definecolor{b2}{RGB}{147,205,221}
\definecolor{b3}{RGB}{75,172,198}
\definecolor{b4}{RGB}{49,133,156}
\definecolor{b5}{RGB}{33,89,104}

\definecolor{g1}{RGB}{216,228,192}
\definecolor{g2}{RGB}{196,214,160}
\definecolor{g3}{RGB}{157,187,97}
\definecolor{g4}{RGB}{121,149,64}
\definecolor{g5}{RGB}{80,99,43}

\definecolor{p1}{RGB}{221,215,230}
\definecolor{p2}{RGB}{205,194,217}
\definecolor{p3}{RGB}{171,154,192}
\definecolor{p4}{RGB}{127,101,159}
\definecolor{p5}{RGB}{84,66,107}

\definecolor{y1}{RGB}{255,230,153}
\definecolor{y2}{RGB}{255,217,102}
\definecolor{y3}{RGB}{255,192,0}
\definecolor{y4}{RGB}{191,144,0}
\definecolor{y5}{RGB}{127,96,0}

\definecolor{c1}{RGB}{251,216,187}
\definecolor{c2}{RGB}{249,196,154}
\definecolor{c3}{RGB}{245,157,86}
\definecolor{c4}{RGB}{234,112,14}
\definecolor{c5}{RGB}{156,75,9}

\newcommand{\btext}[1]{\textcolor{b4}{#1}}

\makeatletter
\algocf@newcommand{KwDataXX}[1]{%
  \sbox\algocf@inputbox{\hbox{\KwSty{Data}\algocf@typo: }}%
  \ifthenelse{\boolean{algocf@inoutnumbered}}{\relax}{\everypar={\relax}}%
  {\let\\\algocf@newinput\hspace{\wd\algocf@inputbox}\hangindent=\wd\algocf@inputbox\hangafter=\wd\algocf@inputbox#1\par}%
  \algocf@linesnumbered
}
\makeatother

\makeatletter
\algocf@newcommand{KwResultXX}[1]{%
  \sbox\algocf@inputbox{\hbox{\KwSty{Result}\algocf@typo: }}%
  \ifthenelse{\boolean{algocf@inoutnumbered}}{\relax}{\everypar={\relax}}%
  {\let\\\algocf@newinput\hspace{\wd\algocf@inputbox}\hangindent=\wd\algocf@inputbox\hangafter=\wd\algocf@inputbox#1\par}%
  \algocf@linesnumbered
}
\makeatother

\usetikzlibrary{decorations.pathreplacing,calc}

\begin{document}

\title{High Bandwidth Memory on FPGAs:\\A Data Analytics Perspective
\\{\normalfont\large
Kaan Kara\IEEEauthorrefmark{1}\IEEEauthorrefmark{2}, Christoph Hagleitner\IEEEauthorrefmark{2}, Dionysios Diamantopoulos\IEEEauthorrefmark{2}, Dimitris Syrivelis\IEEEauthorrefmark{2}, Gustavo Alonso\IEEEauthorrefmark{1}
}\\[-1.5ex]
}
\author{
\IEEEauthorblockA{\IEEEauthorrefmark{1}
Systems Group, Department of Computer Science\\
ETH Zurich, Switzerland, 
firstname.lastname@inf.ethz.ch}
\\\vspace*{-1.5cm}
\and
\IEEEauthorblockA{\IEEEauthorrefmark{2}
IBM Research\\
Zurich, Switzerland}
}

\maketitle

\begin{abstract}
FPGA-based data processing in datacenters is increasing in popularity due to the demands of modern workloads and the ensuing necessity for specialization in hardware.
Driven by this trend, vendors are rapidly adapting reconfigurable devices to suit data and compute intensive workloads.
Inclusion of High Bandwidth Memory (HBM) in FPGA devices is a recent example.
HBM promises overcoming the bandwidth bottleneck, faced often by FPGA-based accelerators due to their throughput oriented design.
In this paper, we study the usage and benefits of HBM on FPGAs from a data analytics perspective.
We consider three workloads that are often performed in analytics oriented databases and implement them on FPGA showing in which cases they benefit from HBM: range selection, hash join, and stochastic gradient descent for linear model training.
We integrate our designs into a columnar database (MonetDB) and show the trade-offs arising from the integration related to data movement and partitioning.
In certain cases, FPGA+HBM based solutions are able to surpass the highest performance provided by either a 2-socket POWER9 system or a 14-core XeonE5 by up to 1.8x (selection), 12.9x (join), and 3.2x (SGD).
\end{abstract}


\section{Introduction}

Performing advanced data analytics efficiently in the datacenter is becoming more important, driven mainly by the exponentially increasing amount of data to be processed and by the rise of machine learning.
Due to the stagnating performance of general purpose processors in recent years, specialized hardware solutions are being considered by major cloud providers as an alternative to achieve high performance and energy efficiency. 
Prominent examples include the development of the Tensor Processing Unit (TPU) by Google to accelerate deep learning~\cite{jouppi2017datacenter}, the usage of FPGAs by Microsoft in their datacenters to offload computation to the network~\cite{putnam2014reconfigurable} and perform low latency inference~\cite{chung2018serving}, the design of interconnects such as OpenCAPI~\cite{stuecheli2018ibm} allowing easy integration of accelerators into systems, and FPGA instances being included in cloud offerings of AWS~\cite{awsf1} and Baidu~\cite{baidu}.

Thanks to their architectural flexibility, FPGAs are widely used for a variety of data processing tasks including machine learning~\cite{fowers2018configurable,umuroglu2017finn,kara2017fpga}, database query processing~\cite{casper2014hardware,kara2017fpga2}, and networking~\cite{istvan2016consensus}.
One of the main bottlenecks faced by current FPGA designs is the bandwidth when accessing data to be consumed or produced.
For throughput-optimized accelerators, that are often implemented on FPGAs, high bandwidth access to data is crucially important.
For instance, recent work on FPGA-based data analytics on partitioning~\cite{kara2017fpga2}, linear model training~\cite{wang2019accelerating}, inference based on decision tree traversal~\cite{owaida2017scalable}, and regular expression matching~\cite{sidler2017accelerating} all mention the bandwidth to memory to be the bottleneck in performance.
Because of this limitation, vendors have started offering FPGA devices with High Bandwidth Memory (HBM).

On Xilinx UltraScale+ devices~\cite{ultrascale}, the HBM exposes a wide bus (8192-bits) to the FPGA fabric, via 32 256-bit AXI3 interfaces. When the logic is clocked at 400 MHz, this bus provides a theoretical peak bandwidth of 410~GB/s towards the HBM with a size of 8~GiB. While this is an improvement over existing systems such as PCIe-attached FPGA cards~\cite{vcu1525} with four DDR4 banks (72 GB/s max), or coherently-attached FPGAs as in Intel Xeon+FPGA~\cite{oliver2011reconfigurable} (20 GB/s max), the architecture raises questions about the usability of the bandwidth in a practical setting.

In this paper, we answer questions that arise regarding the usage of HBM on an FPGA from a data analytics perspective: (1) What are the characteristics of the workloads that benefit most from using the HBM on an FPGA? (2) How does the data partitioning and address space usage affect total usable bandwidth? (3) What does the system look like regarding computation and data movement with an HBM-enabled FPGA when integrating accelerators into a large-scale system such as a database? (4) What FPGA-related challenges such as timing closure and floorplanning arise when using the HBM? 

To answer these questions, we implement three workloads often used in advanced analytics scenarios. The first is range selection, performing a scan of a data column and produing the indexes of values within the specified range. This represents a memory bound operation. The second is a relational join operation, commonly used in databases to match two tables on a given attribute. Joins can be both data and compute intensive, containing also many irregular access patterns~\cite{balkesen2013main}.
The third workload involves training generalized linear models (GLM) using stochastic gradient descent~\cite{bubeck2015convex}, a commonly used model and optimization algorithm in machine learning. This represents a more compute intensive workload with iterative access to the data and frequent updates to its state, that is the model under training.

Our contributions are as follows:

\begin{itemize}[leftmargin=*]
\item We present a system architecture enabling the design of accelerators taking advantange of the HBM on the FPGA while working as part of a database management system.
\item We perform microbenchmarks showing under which circumstances the peak bandwidth of HBM can be achieved.
\item We design and integrate three acceleration solutions into the presented system architecture: range selection, join, and stochastic gradient descent; covering data analytics and in-database machine learning use cases.
\item We evaluate each accelerator solution in-depth showing how and when they benefit from the HBM and compare the resulting performance to high-end multi-core CPUs.
\end{itemize}

\section{Background}
\label{sec:background}

\begin{figure}[t]
\centering
\includegraphics[width=.8\columnwidth]{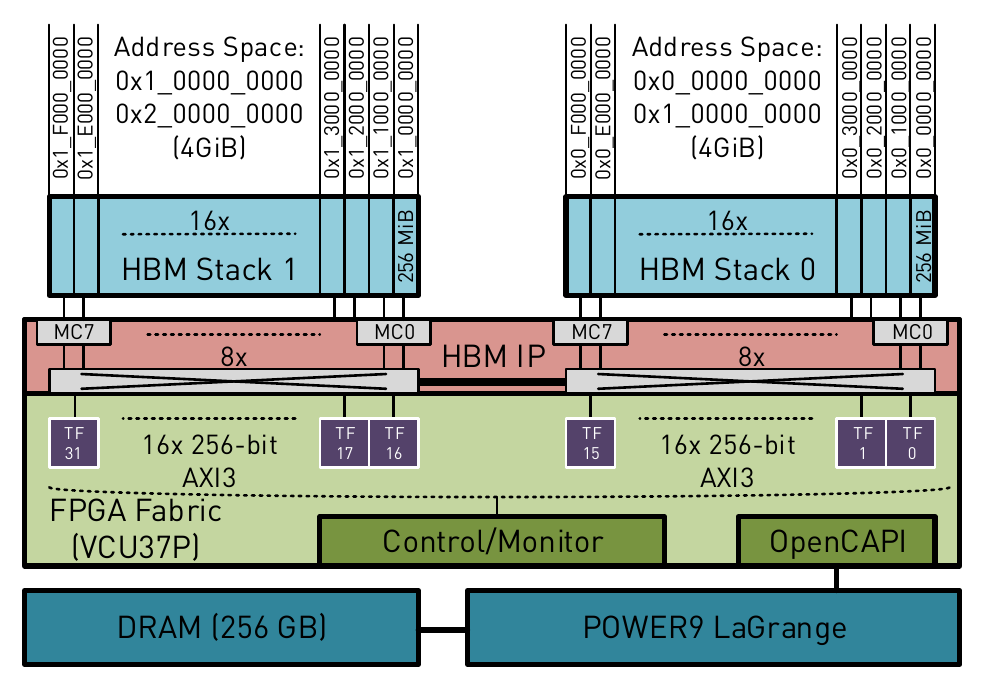}
\caption{An overview for the HBM stacks and the IP, exposing 32 AXI3 interfaces, each 256 bits wide~\cite{hbmdoc}.}
\label{fig:hbm}
\end{figure}

\noindent
\textbf{HBM architecture on Xilinx FPGAs. }
Figure~\ref{fig:hbm} shows an overview of the HBM architecture on Xilinx UltraScale+ FPGAs, based on the documentation for the HBM IP Core~\cite{hbmdoc}.
There are 2 HBM stacks, each with 16 so-called pseudo memory channels that are connected to the FPGA with a 64-bit wide port.
The HBM IP provided by Xilinx contains 16 memory controllers (each resposible for 2 pseudo channels) and a 32$\times$32 crossbar.
The IP converts the 64-bit wide port to a 256-bit wide one, to reduce the clock frequency requirements of the FPGA. Each 256-bit port is exposed to the FPGA fabric as an AXI3 port, in total 32 of them. The crossbar allows each of the AXI3 ports to access the entire address space, 8~GiB in total.
When using the HBM IP, we leave most of the default configurations in the 2 stack configuration, apart from changing the traffic pattern to \textit{linear}, since the workloads of interest will access data in a sequential manner.

When utilizing the crossbar, any congestion on a particular memory channel 
reduces the effective bandwidth.
For instance, if all AXI3 ports try to access the first channel (address space between 0 and 256~MiB), the effective bandwidth is 1/32th of the highest achievable one. We show this behavior later with the microbenchmarks.

\noindent
\textbf{The target platform }
is the Alpha Data ADM-PCIE-9H7~\cite{ad9h7} (AD9H7) card attached to a POWER9 machine via OpenCAPI~\cite{stuecheli2018ibm}, as shown in Figure~\ref{fig:hbm}. The FPGA on AD9H7 is an engineering sample device (Xilinx XCVU37P-2E).
On the FPGA, the control and CPU interfaces are implemented in SystemVerilog. The compute engines discussed later in Sections~\ref{sec:selection},~\ref{sec:join}, and~\ref{sec:sgd} are implemented with Vivado HLS.


\noindent
\textbf{Evalution Setup.} Our baseline CPU experiments are on a 2-socket POWER9 system, each socket with 22 cores at 3.9 GHz.
As an x86 baseline, we use a single-socket Xeon Broadwell E5 with 14 cores~\cite{xeonE5} at 3.5 GHz. On POWER9, we use xlc 16 and compile with \textit{"-O345 -qhot -qaltivec"}. On XeonE5, we use gcc 5.4 and compile with \textit{"-O3 -march=native"}.

\noindent
\textbf{Microbenchmarking HBM on AD9H7. }
We perform microbenchmarks on the FPGA to get an understanding on the behavior of HBM. In particular, we focus on the bandwidth that can be achieved (1) depending on how many ports are utilized and (2) the address space accessed by each port.
The infrastructure for performing these benchmarks are shown in Figure~\ref{fig:hbm}.
Each AXI3 port that the HBM IP exposes is connected to a standalone traffic generator (TG), that can be controlled dynamically by the host. Each TG has 4 configuration parameters: (1) address, (2) size, (3) iterations, (4) read or write.
With these parameters, we can generate traffic individually on each port, that is for instance heavy on sequential reads/writes to measure maximum bandwidth, or heavy on single short accesses to measure latency.

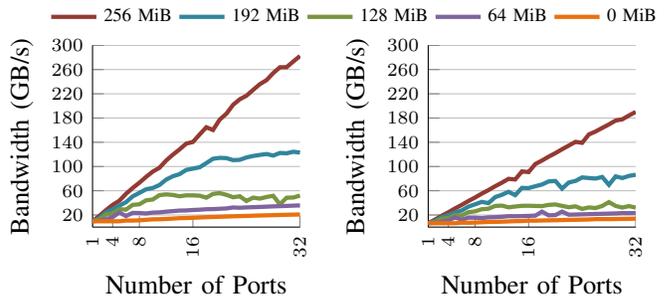
\begin{figure}[t]
\centering
\begin{subfigure}[t]{.49\columnwidth}
\begin{tikzpicture}
\begin{axis}[
    xlabel=Number of Ports,
	ylabel=Bandwidth (GB/s),
    width=\columnwidth,
    height=4cm,
    xmin=1, xmax=32,
    ymin=0, ymax=300,
    axis x line*=bottom,
    axis y line*=left,
    ymajorgrids,
    ytick={20,60,100,140,180,220,260,300},
    y tick label style = {font=\scriptsize},
    xtick={1,4,8,16,32},
    x tick label style = {font=\scriptsize, rotate=90},
    every axis plot/.append style={ultra thick},
    legend columns=5,
    legend style={at={(-0.25,1.05)},anchor=south west,font=\scriptsize,draw=none}
]
\addplot[color=r4,mark=none,] table [x index=0, y index=1] {plotdata/ubenchmark_bw_read_300.dat};
\addplot[color=b4,mark=none] table [x index=0, y index=3] {plotdata/ubenchmark_bw_read_300.dat};
\addplot[color=g4,mark=none] table [x index=0, y index=5] {plotdata/ubenchmark_bw_read_300.dat};
\addplot[color=p4,mark=none] table [x index=0, y index=7] {plotdata/ubenchmark_bw_read_300.dat};
\addplot[color=c4,mark=none] table [x index=0, y index=9] {plotdata/ubenchmark_bw_read_300.dat};
\addlegendentry{256~MiB}
\addlegendentry{192~MiB}
\addlegendentry{128~MiB}
\addlegendentry{64~MiB}
\addlegendentry{0~MiB}
\end{axis}
\end{tikzpicture}
\caption{AXI3 frequency: 300~MHz}
\label{fig:ubenchmark_bw_300}
\end{subfigure}
\begin{subfigure}[t]{.49\columnwidth}
\begin{tikzpicture}
\begin{axis}[
    xlabel=Number of Ports,
	ylabel=Bandwidth (GB/s),
    width=\columnwidth,
    height=4cm,
    xmin=1, xmax=32,
    ymin=0, ymax=300,
    axis x line*=bottom,
    axis y line*=left,
    ymajorgrids,
    ytick={20,60,100,140,180,220,260,300},
    y tick label style = {font=\scriptsize},
    xtick={1,4,8,16,32},
    x tick label style = {font=\scriptsize, rotate=90},
    every axis plot/.append style={ultra thick},
    legend columns=5,
    legend style={at={(-0.15,1.05)},anchor=south west,font=\scriptsize,draw=none}
]
\addplot[color=r4,mark=none,] table [x index=0, y index=1] {plotdata/ubenchmark_bw_read_200.dat};
\addplot[color=b4,mark=none] table [x index=0, y index=3] {plotdata/ubenchmark_bw_read_200.dat};
\addplot[color=g4,mark=none] table [x index=0, y index=5] {plotdata/ubenchmark_bw_read_200.dat};
\addplot[color=p4,mark=none] table [x index=0, y index=7] {plotdata/ubenchmark_bw_read_200.dat};
\addplot[color=c4,mark=none] table [x index=0, y index=9] {plotdata/ubenchmark_bw_read_200.dat};
\end{axis}
\end{tikzpicture}
\caption{AXI3 frequency: 200~MHz}
\label{fig:ubenchmark_bw_200}
\end{subfigure}
\caption{Total read bandwith, depending on the number of ports and address separation per port.
}
\label{fig:ubenchmark_bw}
\end{figure}

Figure~\ref{fig:ubenchmark_bw} shows the read bandwidth measured depending on the number of ports used and address separation per port. The address separation is created by setting the offset of each traffic generator according to the following formula: 
\[
offset = S \times 1~\text{MiB} \times (TF_{id}-1)
\]
where $S=\{256,192,128,64,0\}$ and $TF_{id}=[1,32]$. In the ideal case, when the separation is 256~MiB, we measure 282~GB/s (300~MHz) and 190~GB/s (200~MHz) with 32 ports active. In the worst case, when the separation is 0~MiB (all ports access the same physical HBM channel), we measure 21~GB/s (300~MHz) and 14~GB/s (200~MHz) with 32 ports active.
The experiment when repeated for writes yields very similar results to reads.
The outcome of this benchmark shows the importance of ideally partitioning the data on the HBM to be consumed or produced by the compute engines on the FPGA. \textit{The highest bandwidth benefits can only be observed when each port accesses its own physical memory channel.}

The reasons for not reaching the theoretical bandwidth of 410~GB/s are: (1) We perform the benchmark at highest 300~MHz for AXI3, whereas 400~MHz is required to reach the theoretical maximum.
For scale-out architectures, meeting timing is difficult, mainly due to Super-Logic-Region (SLR) crossings: All HBM ports are connected to SLR0, so any compute engine that cannot be placed at SLR0 has to cross SLRs.
When implementing algorithms later, we found reaching 300~MHz reliably at high utilization is not possible, so we use 200~MHz for all the presented designs for the rest of the paper. This leads to the highest usable bandwidth of 190~GB/s in the ideally partitioned case.
(2) Since our target FPGA is an engineering sample device, the HBM crossbar needs to be clocked at 800~MHz rather than 900~MHz due to a silicon issue, leading to slight decrease in the available bandwidth.


\noindent
\textbf{MonetDB}, a database management system (DBMS),
is used as a baseline for the database workloads on the CPU and to integrate our FPGA-based accelerators. MonetDB~\cite{nes2012monetdb} is a column-oriented in-memory database optimized for online analytical processing (OLAP). An HBM-enhanced acceleration system might improve such a DBMS in particular:
(1) OLAP workloads tend to be read-heavy where data is scanned frequently to extract information via reduction operations such as selection or aggregation.
(2) Column-oriented DBMS perform sequential access frequently and materialize their intermediate results heavily; in both cases, memory bandwidth plays an important role.
(3) The query processing engine of the DBMS is optimized to work on data residing in main memory; compared to disk-optimized databases, this increases the potential of offloading tasks to accelerators.
Recent work has focused on integrating FPGA-based accelerators to MonetDB to improve relational query processing~\cite{kara2017fpga2,sidler2017accelerating} and in-database machine learning~\cite{alonso2019doppiodb,kara2019doppiodb} via a user-defined-function (UDF) interface. In this work, we follow a similar approach, however focusing on the efficient usage of HBM, which to our knowledge is the first system to do so.

\section{System Architecture}
\label{sec:arch}

\begin{figure}[t]
\centering
\includegraphics[width=.8\columnwidth]{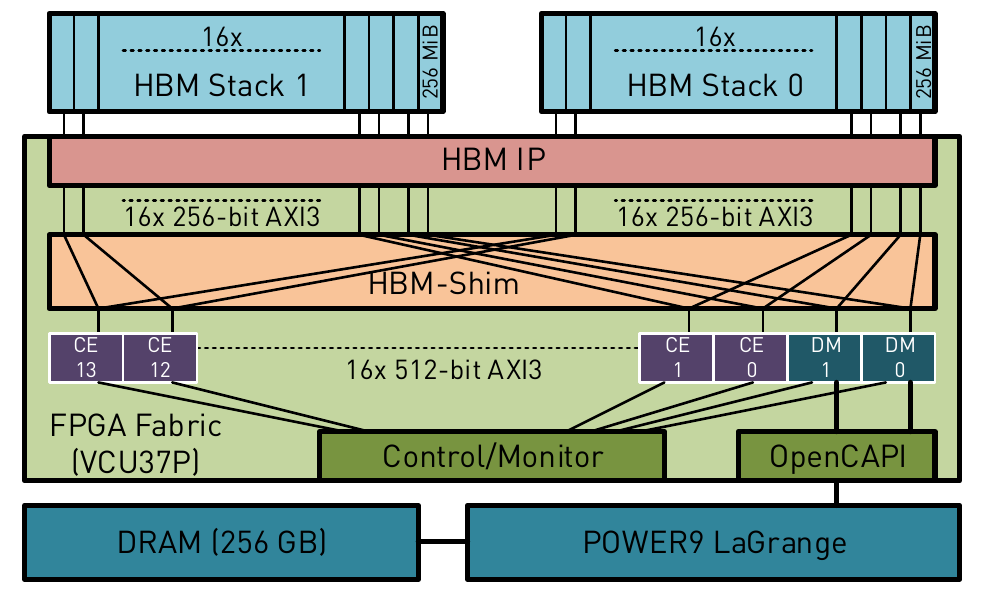}
\caption{An overview of the system architecture.}
\label{fig:system}
\end{figure}

Figure~\ref{fig:system} shows the architecture built on the target platform. Various compute engines (CE) on the FPGA take advantage of HBM and are integrated as operators into MonetDB.
The design decisions behind this architecture are as follows:

\noindent
\textbf{Simplifying HBM Interface.} HBM IP by default offers 32 AXI3 ports, each 256-bit wide (Section~\ref{sec:background}). We showed the highest bandwidth can be reached only if the physical memory channels are accessed in parallel. This can be ensured (1) via data partitioning across the address space during runtime or (2) statically merging ports where a constant offset is applied to some ports to ensure ideal separation.
We use a mix of these: The HBM-shim in Figure~\ref{fig:system} 
uses static port merging for ideal stack partitioning; and otherwise runtime partitioning is utilized when possible.
The shim merges 2 ports from the first and second stacks (ports 0 and 16, 1 and 17, etc.) to form a 512-bit AXI3 port, where a constant offset is applied always to the second port ensuring that there are no inter-stack accesses.
The HBM-shim serves multiple purposes: It simplifies the HBM interface, ensures ideal separation across the two HBM stacks, and reduces the burden of controlling 32 parallel engines by half.

\noindent
\textbf{Data Movement.} In our target system (in-memory DBMS), the data to be consumed resides in the main memory of the CPU.
How to bring this data to the FPGA to be consumed by the CEs which are also attached to the HBM is one of the problems that the presented architecture aims to solve.
A first alternative is to connect each CE both to the DMA (communicating with the CPU) and to one port on HBM. This however (1) creates a bandwidth imbalance and (2) requires arbitration when accessing the DMA by multiple CEs, complicating system architecture and leading to difficult routing.
Instead, we choose a datamover based solution: Two dedicated datamovers are employed to move data between the CPU memory and the HBM. The datamovers occupy 2 of the 16 ports that the HBM-shim exposes.
The remaining ports are usable by CEs, consuming/producing data via the HBM.

\noindent
\textbf{Scale-Out Computation.} Any solution benefiting from the HBM should be a scale-out parallel one. Therefore, we attach multiple CEs to the HBM, each connected to a central control unit, that can asynchronously start/stop and monitor the CEs.
The control unit is exposed to the CPU via a register read/write interface enabling asynchronous control by software:
So, each CE can be started/stopped individually in parallel. Where necessary, synchronization among them (e.g., barriers) can be implemented via software.

\section{Range Selection}
\label{sec:selection}

The first operator we implement based on the system architecture introduced in Section~\ref{sec:arch} is a simple bandwidth-bound and trivially parallel range selection (Algorithm~\ref{algo:selection}).
An array of integers is scanned and each one is checked whether it is within a given range.
If so, its index is put to a result array and the count of matching items is incremented.
Depending on the selectivity of the workload (what proportion of the input matches the range), the runtime for this operation differs: 
For instance, if selectivity is high (all input integers are in the given range), the number of results that need to be written back is larger, increasing the bandwidth requirements.

\begin{algorithm}[t]
\begin{scriptsize}
\KwData{\btext{int} input[num\_items], \btext{int} lower, \btext{int} upper}
\KwResult{\btext{uint} result[num\_items], \btext{uint} num\_matches}
num\_matches = 0\\
\For{i: 0 to num\_items-1}{
    \If{input[i] $>$ lower and input[i] $<$ upper} {
        result[num\_matches++] = i;
    }
}
\end{scriptsize}
\caption{Range selection algorithm, providing the indexes of items that are in the given range.}
\label{algo:selection}
\end{algorithm}

\begin{figure}[t]
\centering
\includegraphics[width=.7\columnwidth]{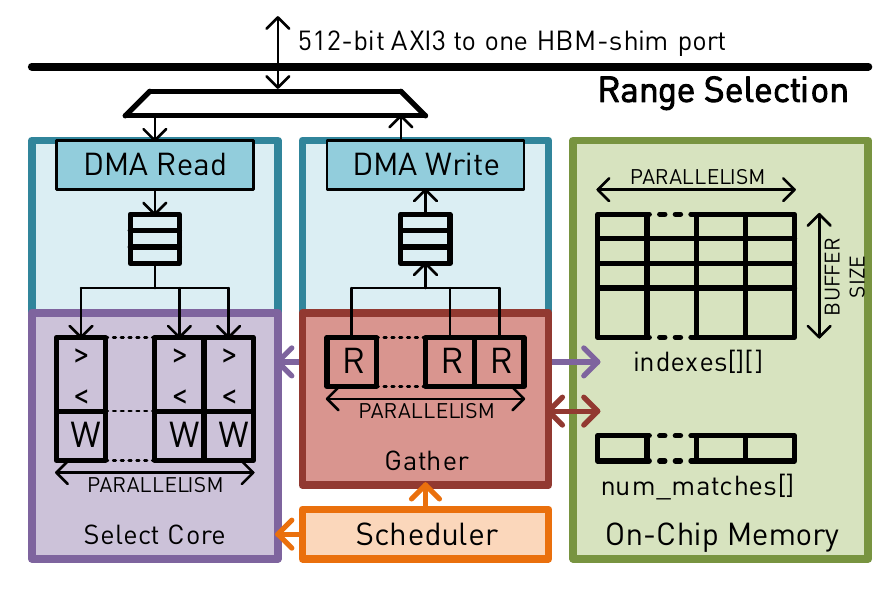}
\caption{Overview of the engine for range selection. \textit{PARALLELISM} equals 16, to be able consume and produce 16 32-bit integers at every cycle and to match the AXI3 port bandwidth.}
\label{fig:selection}
\end{figure}

\begin{figure}[t]
\centering
\begin{subfigure}[t]{.49\columnwidth}
\begin{tikzpicture}
\begin{axis}[
    xlabel=Number of Threads,
	ylabel=Rate (GB/s),
    width=\columnwidth,
    height=3.5cm,
    xmin=0, xmax=256,
    ymin=0, ymax=160,
    axis x line*=bottom,
    axis y line*=left,
    ymajorgrids,
    ytick={20,40,60,80,100,120,140,160},
    y tick label style = {font=\scriptsize},
    xtick={16,64,128,256},
    x tick label style = {font=\scriptsize, rotate=0},
    every axis plot/.append style={ultra thick},
    legend columns=2,
    legend style={at={(0,1.05)},anchor=south west,font=\scriptsize,draw=none}
]
\addplot[color=r4,mark=none,] table [x index=0, y index=2] {plotdata/selection_strong.dat};
\addplot[color=b4,mark=none] table [x index=0, y index=1] {plotdata/selection_strong.dat};
\addplot[color=g4,mark=none] table [x index=0, y index=3] {plotdata/selection_strong.dat};
\addplot[color=c4,mark=none] table [x index=0, y index=4] {plotdata/selection_strong.dat};
\addlegendentry{POWER9}
\addlegendentry{XeonE5}
\addlegendentry{FPGA-partitioned}
\addlegendentry{FPGA-nonpartitioned}
\end{axis}
\draw [thick,<-](0.2, 1.7) -- (0.6, 1.4) node [anchor=west] {14 engines used};
\end{tikzpicture}
\caption{Strong scaling: Number of input items is constant at $128 \cdot 10^6$.}
\label{fig:selection_strong}
\end{subfigure}
\begin{subfigure}[t]{.49\columnwidth}
\begin{tikzpicture}
\begin{axis}[
    xlabel=Number of Threads,
    width=\columnwidth,
    height=3.5cm,
    xmin=0, xmax=256,
    ymin=0, ymax=160,
    axis x line*=bottom,
    axis y line*=left,
    ymajorgrids,
    ytick={20,40,60,80,100,120,140,160},
    y tick label style = {font=\scriptsize},
    xtick={16,64,128,256},
    x tick label style = {font=\scriptsize, rotate=0},
    every axis plot/.append style={ultra thick},
    legend columns=2,
    legend style={at={(0,1.05)},anchor=south west,font=\scriptsize,draw=none}
]
\addplot[color=r4,mark=none] table [x index=0, y index=2] {plotdata/selection_weak.dat};
\addplot[color=b4,mark=none] table [x index=0, y index=1] {plotdata/selection_weak.dat};
\addplot[color=g4,mark=none] table [x index=0, y index=3] {plotdata/selection_weak.dat};
\addplot[color=c4,mark=none] table [x index=0, y index=4] {plotdata/selection_weak.dat};
\end{axis}
\draw [thick,<-](0.2, 1.75) -- (0.6, 1.45) node [anchor=west] {14 engines used};
\end{tikzpicture}
\caption{Weak scaling: Base ($16 \cdot 10^6$) is multiplied with num. threads.}
\label{fig:selection_weak}
\end{subfigure}
\caption{Processing rate for range selection depending on the number of threads on different platforms.
The selectivity is 0\%, so no item matches the given range.}
\label{fig:selection_perf}
\end{figure}
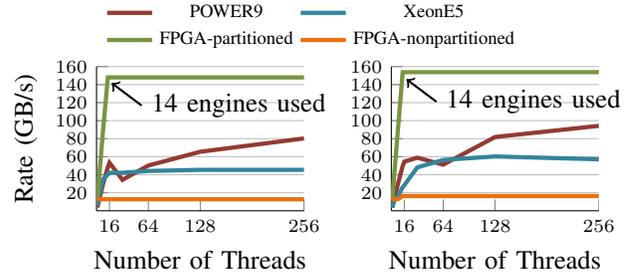
\begin{figure}[t]
\begin{tikzpicture}
\begin{axis}[
    xlabel=Selectivity (\%),
	ylabel=Rate (GB/s),
	xmode=log,
	log ticks with fixed point,
    width=\columnwidth,
    height=3.5cm,
    xmin=0.001, xmax=100,
    ymin=0, ymax=160,
    axis x line*=bottom,
    axis y line*=left,
    ymajorgrids,
    ytick={20,40,60,80,100,120,140,160},
    y tick label style = {font=\scriptsize},
    xtick=data,
    x tick label style = {font=\scriptsize, rotate=0},
    xtick={0.001,0.01,0.1,1,10,20,40,100},
    every axis plot/.append style={ultra thick},
    legend columns=2,
    legend style={at={(0,1.05)},anchor=south west,font=\scriptsize,draw=none}
]
\addplot[color=r4,mark=none,] table [x index=0, y index=2] {plotdata/selectivity_effect.dat};
\addplot[color=b4,mark=none] table [x index=0, y index=1] {plotdata/selectivity_effect.dat};
\addplot[color=g4,mark=none] table [x index=0, y index=3] {plotdata/selectivity_effect.dat};
\addplot[color=g2,mark=none] table [x index=0, y index=4] {plotdata/selectivity_effect.dat};
\addplot[color=c4,mark=none] table [x index=0, y index=5] {plotdata/selectivity_effect.dat};
\addplot[color=c2,mark=none] table [x index=0, y index=6] {plotdata/selectivity_effect.dat};
\addlegendentry{POWER9-256 threads}
\addlegendentry{XeonE5-64 threads}
\addlegendentry{FPGA-partitioned-14}
\addlegendentry{FPGA-partitioned-14 (copy)}
\addlegendentry{FPGA-nonpartitioned-14}
\addlegendentry{FPGA-nonpartitioned-14 (copy)}
\end{axis}
\end{tikzpicture}
\caption{Effect of selectivity. With higher selectivity, more output is generated.
The FPGA numbers with \textit{(copy)} include the copying time of output data from the FPGA to the CPU.}
\label{fig:selectivity}
\end{figure}
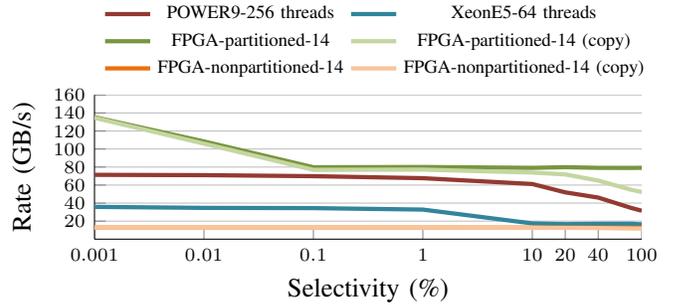

\noindent
\textbf{FPGA Implementation.}
Figure~\ref{fig:selection} shows the engine performing range selection on the FPGA.
There are 2 main pipelines (ingress and egress) that are activated one after the other by the scheduler. The granularity of switching between ingress/egress is determined by the \textit{BUFFER SIZE} (1024). With \textit{PARALLELISM} set to 16 (to match the 512-bit wide AXI3 port bandwidth), this results in an internal buffer size of 64~KiB to store resulting indexes.

At the ingress pipeline, a \textit{DMA read} module reads data from the HBM using the 512-bit AXI3 port and pushes it into a 512-bit wide FIFO.
The FIFO is read by the \textit{Select Core}, that internally has 16 parallel comparison and update units performing the selection (Algorithm~\ref{algo:selection}, lines 3-5).
The resulting indexes and number of matches can be written to on-chip memory (Block-RAM or Ultra-RAM) in parallel by the update units thanks to the spatial partitioning.

At the egress pipeline, the results stored in on-chip memory are read in parallel, a 512-bit line is constructed and written back to HBM via \textit{DMA Write}.
At this stage, since the number of matches from 16 individual units might differ, we append a dummy element to the line whenever necessary, when writing the results back.
In the end, this causes some more data to be written than necessary; the same trick needs to be employed also on the CPU when utilizing SIMD instructions.

\noindent
\textbf{Evaluation.}
We use \textit{one bitstream} with 14 selection engines utilizing all 14 AXI3 ports exposed by the HBM-shim (Figure~\ref{fig:system}).
The number of engines used and the data placement are runtime decisions.
Figure~\ref{fig:selection_perf} shows the processing rate with strong/weak scaling, over the number of threads used.
The selectivity is 0\%, so no output is generated and the focus is on the data consumption rate.
\textit{For the selection operator only} we assume the input data is already in the HBM.
This is reasonable for bandwidth sensitive operators, because the DBMS during the initial execution of a query brings the data from the disk: The first query takes much longer than subsequent ones, similar to bringing data to HBM for the first query, which in this case is more expensive than performing selection since OpenCAPI bandwidth is lower than HBM.

When input data is partitioned in an ideal way, such that the engines access data from their own physical memory channel, the selection can be performed at 154 GB/s with 14 engines: 11 GB/s per engine, where theoretical maximum is 12.8 GB/s.
Even in the weak scaling case, when XeonE5 reaches its memory bandwidth, the FPGA can outperform XeonE5 by 2.7x (57 GB/s with 256 threads) and POWER9 by 1.6x (94 GB/s with 256 threads).
When the data is not partitioned however, the advantage of HBM is diminished and processing rate with all 14 engines drops to 16~GB/s.

Figure~\ref{fig:selectivity} shows the effect of selectivity: With increased selectivity, the output generated by the range selection increases. When selectivity is 100\%, the size of the generated output is equal to the size of the input. This has a diminishing effect on the input consumption rate of the FPGA, dropping to 80~GB/s, because the AXI3 port needs to be shared between reads and writes to the HBM.
We also show how the rate drops when we include the copying time of output data from FPGA to the CPU. This is necessary in case the output is to be consumed further by the DBMS.
The copying has little effect for low selectivity, however is increasingly more important for high selectivity since more output is produced. 
The same effect is also visible with the CPU experiments.

\section{Join}
\label{sec:join}

In relational databases, joins are frequently performed especially in analytical settings.
There is a large body of work optimizing join algorithms for CPUs~\cite{balkesen2013main}, GPUs~\cite{kaldewey2012gpu}, and FPGAs~\cite{kara2017fpga2}.
In this work, we are interested to provide an end-to-end join implementation that can directly be used by the DBMS (MonetDB), while benefiting from HBM.
Much of the related work on join optimization~\cite{balkesen2013main} focuses on finding the number of matching tuples, without materializing (writing back the resulting tuples), which can be expensive but necessary if the join is to be used in a DBMS.
Therefore, we also include the materialization step in our implementation.

\begin{algorithm}[t]
\DontPrintSemicolon
\begin{scriptsize}
\KwData{\btext{int} S[S\_num], \btext{int} L[L\_num] \tcp*{small, large}}
\KwDataXX{num\_threads}
\KwResult{\btext{uint} S\_out[num\_matches], \btext{uint} L\_out[num\_matches]}
\tcc{Naively partition L}
ps = L\_num/num\_threads \tcp*{partition size}
\For{t: 0 to num\_threads-1}{
LP[t][(t+1)*ps-1 ... t*ps] $\leftarrow$ L[(t+1)*ps-1 ... t*ps]
}
HT\_S $\leftarrow$ BuildHT(S) \tcp*{Build a hash table on S}
\tcc{Perform join in parallel for each partition}
\For{t: 0 to num\_threads-1}{
    \For{i: 0 to ps-1}{
        h = hash(LP[t][i])\\
        \While{s = ProbeHT(HT\_S, h) is success}{
            \If{LP[t][i] == s} {
                S\_out[num\_matches] = s\\
                L\_out[num\_matches++] = LP[t][i]
            }
        }
    }
}
\end{scriptsize}
\caption{Naively partitioned hash join.}
\label{algo:join}
\end{algorithm}

MonetDB by default uses a naively partitioned hash join implementation (Algorithm~\ref{algo:join}).
The larger side of the join is partitioned across the number of threads to be used.
The smaller side is used to create one hash table (line 5), that is used by all threads during probing (line 9).
One drawback is that if the hash table does not fit the cache, the latency bound operation of probing the hash table becomes a bottleneck.
More sophisticated algorithms such as radix join~\cite{boncz1999database} perform high fanout radix partitioning to eliminate this problem, making sure the hash table fits the cache.
Recently, Kara et al.~\cite{kara2017fpga2} have shown that high fanout radix partitioning can be accelerated by an FPGA. But their method is useful only if the result of the partitioning is to be consumed by a CPU; so HBM does not provide an advantage for performing this operation, as the results need to be written back to CPU's memory, making the CPU-FPGA interconnect the bottleneck.
Furthermore, in many real world use cases or benchmarks emulating these (such as TPC-H~\cite{boncz2013tpc}), one side of the join is usually much smaller because of a previously performed selection; so radix partitioning becomes less critical.

For these reasons and to also ensure seamless integration with MonetDB, we decide to implement the same hash join algorithm as used by MonetDB (Algorithm~\ref{algo:join}).
This algorithm also fits well with the scale-out processing we look for to benefit as much as possible from the HBM.

\begin{figure}[t]
\centering
\includegraphics[width=.7\columnwidth]{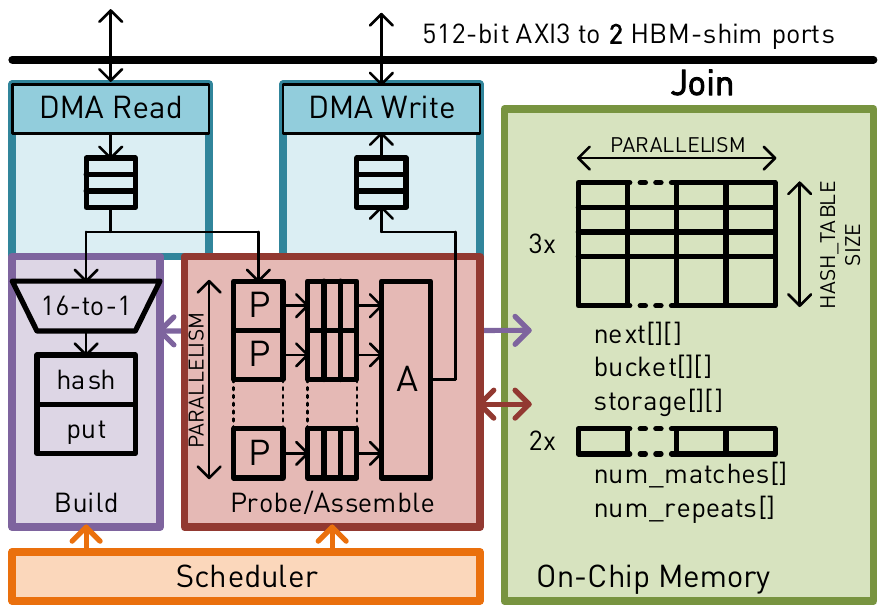}
\caption{Overview of the engine for performing join. \textit{PARALLELISM} equals 16, to be able consume and produce 16 32-bit integers at every cycle and to match the AXI3 port bandwidth.
}
\label{fig:join}
\end{figure}

\noindent
\textbf{FPGA Implementation.}
An overview for the implementation on the FPGA is shown in Figure~\ref{fig:join}.
We aim to optimize the probing step as much as possible, since building the hash table takes a negligible time (uses the smaller side of the join).
Also, parallelizing the build step with SIMD is not possible, since each new insertion depends on the previous one due to possible collisions.
For this reason, the build step is not parallelized and processes items one after the other (a 16-to-1 multiplexer is inserted in \textit{Build} module to reduce the 512-bit line). 
Since we would like to achieve maximum SIMD parallelism during probe, 16 replicas for the same hash table are created, using Ultra-RAM resources.
This is necessary, because during probing each item will point to a different location in the hash table and we would like to obtain 16 probe results in parallel at the same clock cycle.

Once the build step is complete, the probe is started by the scheduler.
\textit{DMA Read}, \textit{Probe/Assemble}, and \textit{DMA Write} all work in a parallel dataflow fashion. Thanks to the 16 replicas of the hash tables created, 16 probes are performed in parallel. The results are assembled at the same rate. If the number of results produced by each of the 16 pipelines differ, a dummy element is added to the 512-bit line during the assemble step at the locations where fewer results are produced.
The entire probe step can thus be performed with an initiation interval (II) of 1, able to consume and produce 512-bits per cycle.
Since each join engine requires 2 AXI3 ports (reading/writing simultaneously), we employ 7 join engines in our system (Figure~\ref{fig:system}) and generate \textit{one bitstream}, utilizing all HBM ports.

\begin{table}[t]
\centering
\caption{Join processing rate with changing input properties and configurations. Number of tuples for the larger side is 512~Million (2~GB) and for the smaller side 4096 (16~KB).}
\label{tab:join}
\begin{scriptsize}
\begin{tabular}{c|c|c|c|c|c|c}
L & S & L & HT\_S & Handle & 1 engine & 7 engines\\
Unique & Unique & Load & Build & Col. & (GB/s) & (GB/s)\\
\hline
 1 & 1 & 1 & 1 & 1 & 1.81 & 6.48\\
 1 & 1 & 0 & 1 & 1 & 2.13 & 14.68\\
 1 & 1 & 1 & 1 & 0 & 6.07 & 10.25\\
 1 & 1 & 0 & 1 & 0 & 12.77 & \textbf{80.95}\\
 1 & 0 & 1 & 1 & 1 & 1.61 & 6.09\\
 1 & 0 & 0 & 1 & 1 & 1.86 & 12.79\\
\end{tabular}
\end{scriptsize}
\end{table}

\noindent
\textbf{Evaluation: Configurations.} Table~\ref{tab:join} shows the processing rate (size of L divided by the runtime) for the end-to-end join, including the copying time of the results from the HBM to the CPU's memory.
For differing configurations we obtain differing processing rates.
The main factors affecting the processing rate are: (1) Loading L from CPU's memory to the HBM and (2) necessity to handle collissions due to S not being unique.
The first factor is less critical: The DBMS also has to load data from disk when a query is executed initially, so subsequent ones are much faster. As long as the larger side of the join fits the HBM, the same can be assumed also for the join on the FPGA.
The second factor is more critical: If S contains non-unique items, collisions need to be handled during probing. Due to the non-deterministic nature of collision handling,  II=1 cannot be upheld leading to reduced processing rate.
So, only joins that have a unique S benefit from HBM fully, reaching 81~GB/s when using 7 engines. An advantage is that this is a frequent case in relational processing, especially for primary-foreign key joins that take place often in real world workloads and benchmarks~\cite{boncz2013tpc}.

\begin{figure}
\begin{subfigure}[t]{\columnwidth}
\begin{tikzpicture}
\begin{axis}[
    xlabel=Number of Threads,
	ylabel=Rate (GB/s),
	ymode=log,
	log ticks with fixed point,
    width=\columnwidth,
    height=3.5cm,
    xmin=0, xmax=64,
    ymin=0.1, ymax=90,
    axis x line*=bottom,
    axis y line*=left,
    ymajorgrids,
    ytick={0.1,0.2,0.4,0.8,1.6,3.2,5,10,20,40,80},
    y tick label style = {font=\scriptsize},
    xtick=data,
    x tick label style = {font=\scriptsize, rotate=0},
    every axis plot/.append style={ultra thick},
    legend columns=2,
    legend cell align={left},
    legend style={at={(-0.1,1.05)},anchor=south west,font=\scriptsize,draw=none}
]
\addplot[color=r4,mark=none] table [x index=0, y index=3] {plotdata/join_strong.dat};
\addplot[color=r3,mark=none,dashed] table [x index=0, y index=4] {plotdata/join_strong.dat};
\addplot[color=b4,mark=none] table [x index=0, y index=1] {plotdata/join_strong.dat};
\addplot[color=b3,mark=none,dashed] table [x index=0, y index=2] {plotdata/join_strong.dat};
\addplot[color=g4,mark=none] table [x index=0, y index=5] {plotdata/join_strong.dat};
\addplot[color=g3,mark=none,dashed] table [x index=0, y index=6] {plotdata/join_strong.dat};
\addplot[color=c4,mark=none] table [x index=0, y index=7] {plotdata/join_strong.dat};
\addplot[color=c3,mark=none,dashed] table [x index=0, y index=8] {plotdata/join_strong.dat};
\addlegendentry{POWER9 (S Unique)}
\addlegendentry{POWER9 (S not-Unique)}
\addlegendentry{XeonE5 (S Unique)}
\addlegendentry{XeonE5 (S not-Unique)}
\addlegendentry{FPGA (L Load, S Unique)}
\addlegendentry{FPGA (L Load, S not-Unique)}
\addlegendentry{FPGA (L not-Load, S Unique)}
\addlegendentry{FPGA (L not-Load, S not-Unique)}
\end{axis}
\end{tikzpicture}
\caption{Join processing rate ($sizeof(L)/runtime$) over number of threads. The number of tuples of L is 512 Million and for S 4096.}
\label{fig:join_strong}
\end{subfigure}
\begin{subfigure}[t]{\columnwidth}
\begin{tikzpicture}
\begin{axis}[
    xlabel=Number of Items in S ($\times$ 1000),
	ylabel=Runtime (s),
	xmode=log,
	log ticks with fixed point,
    width=\columnwidth,
    height=4cm,
    xmin=6, xmax=16000,
    ymin=0, ymax=4,
    axis x line*=bottom,
    axis y line*=left,
    ymajorgrids,
    ytick={0.5,1,1.5,2,2.5,3,3.5,4},
    y tick label style = {font=\scriptsize},
    xtick=data,
    x tick label style = {font=\scriptsize, rotate=45},
    every axis plot/.append style={ultra thick},
    legend columns=2,
    legend cell align={left},
    legend style={at={(-0.1,1.05)},anchor=south west,font=\scriptsize,draw=none}
]
\addplot[color=r4,mark=none] table [x index=0, y index=3] {plotdata/join_runtime.dat};
\addplot[color=r3,mark=none,dashed] table [x index=0, y index=4] {plotdata/join_runtime.dat};
\addplot[color=b4,mark=none] table [x index=0, y index=1] {plotdata/join_runtime.dat};
\addplot[color=b3,mark=none,dashed] table [x index=0, y index=2] {plotdata/join_runtime.dat};
\addplot[color=g4,mark=none] table [x index=0, y index=5] {plotdata/join_runtime.dat};
\addplot[color=g3,mark=none,dashed] table [x index=0, y index=6] {plotdata/join_runtime.dat};
\addplot[color=c4,mark=none] table [x index=0, y index=7] {plotdata/join_runtime.dat};
\addplot[color=c3,mark=none,dashed] table [x index=0, y index=8] {plotdata/join_runtime.dat};
\end{axis}
\end{tikzpicture}
\caption{End-to-end join runtime over the size of S.
\textbf{64 threads} are used for XeonE5 and POWER9.
\textbf{7 engines} are used on the FPGA.
}
\label{fig:join_runtime}
\end{subfigure}
\caption{Join evaluation.}
\label{fig:join_perf}
\end{figure}
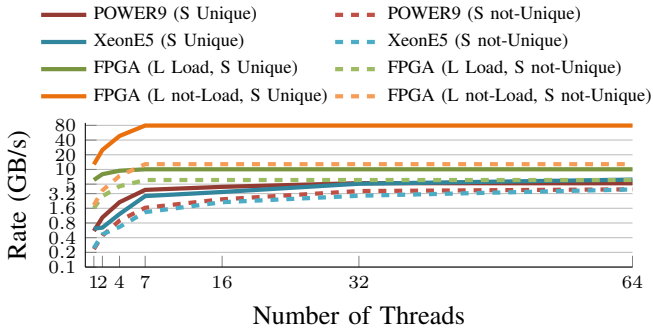
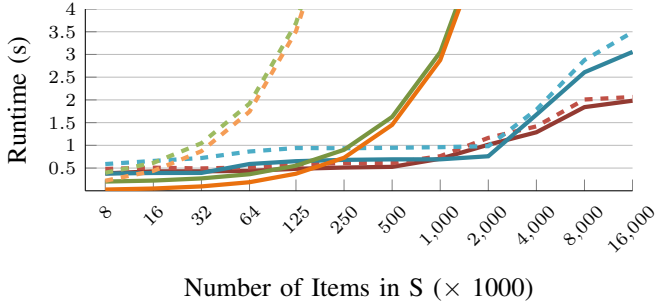

\noindent
\textbf{Evaluation: Scaling.} Figure~\ref{fig:join_strong} shows how the processing rate scales with increasing number of threads. For the CPU baselines, we use MonetDB and measure the time for the individual join operator within the DBMS.
We observe that even for the worst case (when L needs to be loaded and S is not unique so collision handling is required), the FPGA outperforms both XeonE5 and POWER9 using 64 threads. For the ideal case when L is already in the HBM and S is unique, the FPGA can outperform the best rate of XeonE5 by 12.8x.
This rate is achieved when L is written to the HBM such that all engines access their own physical memory channels to benefit most from the HBM.
As shown in Sections~\ref{sec:background} and~\ref{sec:selection}, if this is not done, the high bandwidth advantage is lost.

\noindent
\textbf{Evaluation: Size of the smaller side (S).} In Figure~\ref{fig:join_runtime}, we show the effect of increasing the size of S on the end-to-end join runtime.
This causes a linear increase (log x-axis) for the runtime on the FPGA, whereas a sublinear increase on the CPU as long as S fits the cache.
The main reason for this is our probe-optimized join implementation on the FPGA.
Since we want to benefit from HBM as much as possible during probing, we replicate the hash table 16 times per engine (7 engines in total) using large amount of on-chip memory. This causes the hash table size to be limited to 8192 tuples (16~KiB). When the smaller side is larger than this, we have to repeatedly scan the entire L, causing the linear increase in Figure~\ref{fig:join_runtime}.
The FPGA outperforms the CPU when L is already in HBM and S is unique until S has 125,000 tuples.
From a DBMS perspective this is acceptable: Query optimization and algorithm selection is performed regularly for each query~\cite{nes2012monetdb}.
The FPGA adds an attractive alternative join implementation to the DBMS, that is to be used whenever the smaller side has less than 125,000 tuples, frequent in analytics workloads.

\section{Stochastic Gradient Descent (SGD)}
\label{sec:sgd}

\begin{algorithm}[t]
\DontPrintSemicolon
\begin{scriptsize}
\KwData{$n$, $m$, $\mathbf{x}$[$n$] $\leftarrow$ 0 \tcp*{num\_features, num\_samples, model}}
\KwDataXX{step size $\alpha$, minibatch size $B$}
\KwDataXX{$S(\mathbf{z}) =
\begin{cases} 
\mathbf{z} &\text{for Ridgereg} \\
1/(1+\exp(-\mathbf{z})) &\text{for LogregL2}
\end{cases}$}
\KwResult{\btext{float} $\mathbf{x}$[$n$] \tcp*{trained model}}
\For{epoch: 1 to N} {
    $\mathbf{g} \leftarrow 0$\\
    \For{i: 1 to m}{
        $ dot = S(\langle \mathbf{x}, \mathbf{a}_i \rangle)$ \tcp*{Dot}
        $ dot = \alpha(dot - b_i)$ \tcp*{ScalarEngine}
        $\mathbf{g}$ += $dot \cdot \mathbf{a}^T_i$ \tcp*{Update}
        \If{i mod B is 0} {
            $\mathbf{x}$ = $\mathbf{x} - \alpha(\mathbf{g} + 2\lambda\mathbf{x})$ \tcp*{Update}
            $\mathbf{g} \leftarrow 0$
        }
    }
}
\end{scriptsize}
\caption{SGD implemnetation.}
\label{algo:sgd}
\end{algorithm}

Performing machine learning (ML) within a DBMS is becoming increasingly important, with all major vendors now offering this functionality~\cite{db2ml,oracleml, macgregor2013predictive}.
Since ML has different requirements than traditional analytics, recent work has focused on providing accelerators to perform in-database ML efficiently~\cite{wang2019accelerating,mahajan2018rdbms,kara2018columnml}. In this paper, we focus on training generalized linear models (e.g., linear/logistic regression) with SGD. Recent work shows the advantage of using FPGAs~\cite{kara2017fpga}: The SGD accelerator can be designed as a fully pipelined dataflow architecture, saturating the memory bandwidth (6.5~GB/s) on their target platform (Xeon+FPGA~\cite{oliver2011reconfigurable}). HBM has therefore a large potential for improving SGD with a scale-out architecture on the FPGA and provide an even more pronounced advantage over general purpose processors.

SGD is an optimization algorithm that minimizes problems of the following form:
\begin{equation}
\min_{\mathbf{x} \in \mathbb{R}^n} \left(\frac{1}{m} \sum_{i=1}^m J( \langle \mathbf{x},\mathbf{a}_i \rangle, b_i )\right) +
\underbrace{\lambda \lVert \mathbf{x} \rVert_{2}^{2}}_{\text{For L2-reg}}
\end{equation}
\[
\begin{aligned}
& J =
\begin{cases} 
\frac{1}{2}(\langle \mathbf{x},\mathbf{a}_i \rangle - b_i)^2 \hspace{4cm} \text{for Rigdereg} \\
- b_i \log(h_{\mathbf{x}}(\mathbf{a}_i)) - (1-b_i) \log(1-h_{\mathbf{x}}(\mathbf{a}_i))  \, \text{for Logreg}\\
\end{cases}\\
& h_{\mathbf{x}}(\mathbf{a}_i) = 1/(1+\exp(-\langle \mathbf{x},\mathbf{a}_i \rangle)) \text{ is the sigmoid function.}
\end{aligned}
\label{eq:losses}
\]
where $(\mathbf{a}_1, b_1), ..., (\mathbf{a}_m, b_m) \in ([-1, 1]^n \times \mathbb{R})$ is a set of samples and $J: \mathbb{R}^n \times \mathbb{R} \rightarrow [0, \infty)$ is a non-negative convex loss function.
SGD is shown in Algorithm~\ref{algo:sgd}.

\begin{figure}[t]
\centering
\includegraphics[width=.9\columnwidth]{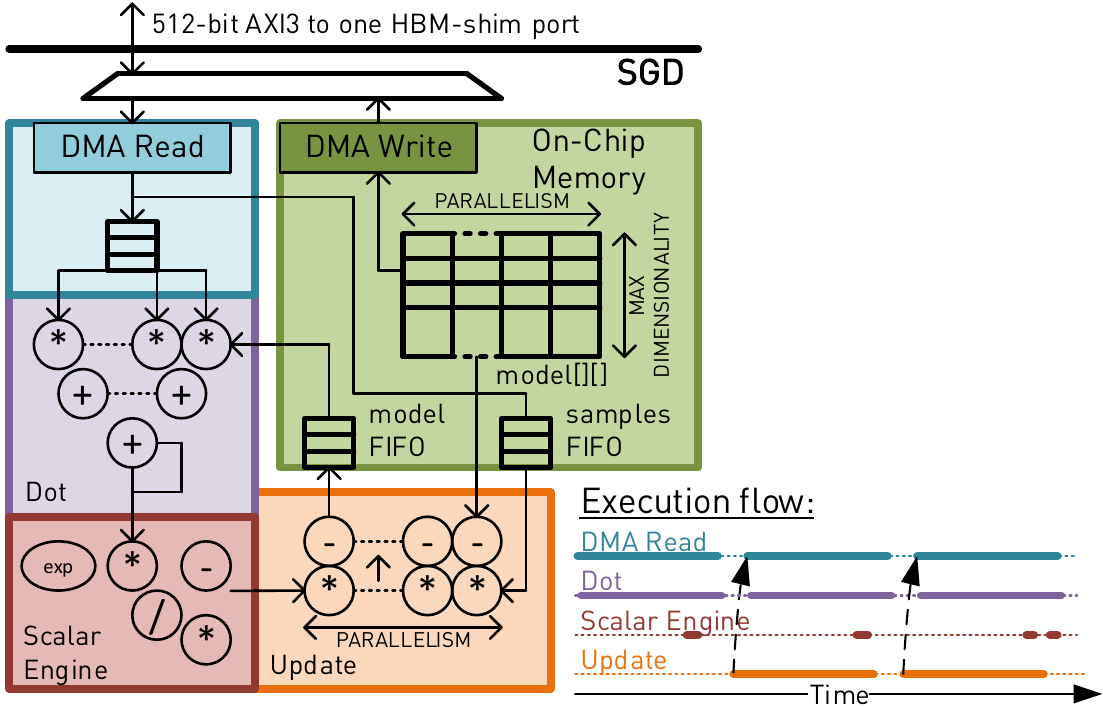}
\caption{Overview of the engine for performing SGD. \textit{PARALLELISM} equals 16, to be able consume and produce 16 32-bit floats at every cycle and to match the AXI3 port bandwidth.
}
\label{fig:sgd}
\end{figure}

\noindent
\textbf{FPGA Implementation. }
We implement FPGA-based SGD following a similar architecture as related work~\cite{kara2017fpga}, shown in Figure~\ref{fig:sgd}.
The dataset is scanned $N$ times (number of epochs) and forwarded to the compute modules.
These modules (\textit{Dot}, \textit{ScalarEngine}, and \textit{Update}) work in a parallel dataflow fashion, as illustrated in Figure~\ref{fig:sgd}, ensuring full utilization.

Unlike Kara et al.~\cite{kara2017fpga} we do not allow stale updates to the model that can happen due to the latency of the entire pipeline and read-after-write dependencies (between lines 4 and 7).
We thus ensure high quality convergence, but have to accept a lower processing rate for low-dimensional datasets and lower minibatch sizes, since the entire pipeline as shown in Figure~\ref{fig:sgd} cannot be filled in those cases.
Nevertheless, considering the processing rate of \textit{only one engine}, even in the worst case we match Kara et al.~\cite{kara2017fpga} (6.5 GB/s) and exceed it by 1.7x in the best case, thanks to the higher bandwidth we get from one HBM-shim port.
We use \textit{one bitstream} with 14 SGD engines in the system architecture (Figure~\ref{fig:system}); all engines capable of performing SGD in parallel, reading data from the HBM and writing the trained model back.

\noindent
\textbf{Evaluation. }
We evaluate the SGD implementation using the datasets listed in Table~\ref{tab:datasets}, considering the popular hyperparameter search use case occurring frequently in real world scenarios~\cite{bergstra2012random}: Multiple models are trained on the same dataset, however with differing hyperparameters, to find the best set that achieves the highest test score.
This problem is trivially parallel, but requires many processors and is usually performed using entire clusters.
Our goal is to show that 1 FPGA with an HBM can replace multiple CPUs for performing this work, in an in-database ML setting.
SGD is an iterative algorithm: data is read multiple times; so the initial copy cost from the CPU to the FPGA is negligible ($<$1\% of total runtime).

\begin{table}[t]
\centering
\caption{Datasets used during evaluation}
\label{tab:datasets}
\setlength\tabcolsep{1pt}
\begin{scriptsize}
\begin{tabular}{c|c|c|c|c|c}
Name & \# Samples & \# Features & \# Classes & \# Epochs & Size (MB)\\
\hline
IM & 41600 & 2048 & binary & 10 & 340.8\\
MNIST & 50000 & 784 & 10 & 10 & 156.8\\
AEA & 32768 & 126 & binary & 20 & 16.5\\
SYN & 262144 & 256 & regression & 10 & 268.435
\end{tabular}
\end{scriptsize}
\end{table}

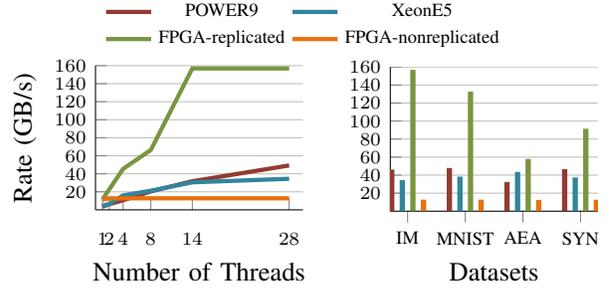
\begin{figure}
\begin{subfigure}[t]{.49\columnwidth}
\begin{tikzpicture}
\begin{axis}[
    xlabel=Number of Threads,
	ylabel=Rate (GB/s),
    width=\columnwidth,
    height=3.5cm,
    xmin=0, xmax=30,
    ymin=0, ymax=160,
    axis x line*=bottom,
    axis y line*=left,
    ymajorgrids,
    ytick={20,40,60,80,100,120,140,160},
    y tick label style = {font=\scriptsize},
    xtick=data,
    x tick label style = {font=\scriptsize, rotate=0},
    xticklabel style={yshift=-5pt},
    every axis plot/.append style={ultra thick},
    legend columns=2,
    legend style={at={(0,1.05)},anchor=south west,font=\scriptsize,draw=none}
]
\addplot[color=r4,mark=none] table [x index=0, y index=2] {plotdata/sgd_rate.dat};
\addplot[color=b4,mark=none] table [x index=0, y index=1] {plotdata/sgd_rate.dat};
\addplot[color=g4,mark=none] table [x index=0, y index=3] {plotdata/sgd_rate.dat};
\addplot[color=c4,mark=none] table [x index=0, y index=4] {plotdata/sgd_rate.dat};
\addlegendentry{POWER9}
\addlegendentry{XeonE5}
\addlegendentry{FPGA-replicated}
\addlegendentry{FPGA-nonreplicated}
\end{axis}
\end{tikzpicture}
\caption{Over number of threads }
\label{fig:sgd_rate}
\end{subfigure}
\begin{subfigure}[t]{.49\columnwidth}
\begin{tikzpicture}
\begin{axis}[
    ybar,
    bar width=2pt,
    xlabel=Datasets,
    width=\columnwidth,
    height=3.5cm,
    ymin=0, ymax=160,
    axis x line*=bottom,
    axis y line*=left,
    ymajorgrids,
    ytick={20,40,60,80,100,120,140,160},
    y tick label style = {font=\scriptsize},
    symbolic x  coords={IM,MNIST,AEA,SYN},
    xtick =data,
    x tick label style = {font=\scriptsize, rotate=0},
    legend columns=2,
    legend style={at={(0,1.05)},anchor=south west,font=\scriptsize,draw=none}
]
\addplot[fill=r4,mark=none,draw=none] table [x index=0, y index=2] {plotdata/sgd_datasets.dat};
\addplot[fill=b4,mark=none,draw=none] table [x index=0, y index=1] {plotdata/sgd_datasets.dat};
\addplot[fill=g4,mark=none,draw=none] table [x index=0, y index=3] {plotdata/sgd_datasets.dat};
\addplot[fill=c4,mark=none,draw=none] table [x index=0, y index=4] {plotdata/sgd_datasets.dat};
\end{axis}
\end{tikzpicture}
\caption{Over datasets (28 threads)}
\label{fig:sgd_datasets}
\end{subfigure}
\caption{SGD processing rate. FPGA-replicated shows when the dataset to be consumed is replicated so that each engine accesses it via its own physical memory channel on the HBM.}
\label{fig:sgd_perf}
\end{figure}

\noindent
\textbf{Evaluation: Processing rate.} Figure~\ref{fig:sgd_rate} shows how the processing rate scales with the number of parallel jobs, for training the last layer of an InceptionV3 neural network~\cite{szegedy2016rethinking} for a binary classification problem, using the \textit{IM} dataset. There are 28 hyperparameter configurations to search, resulting in 28 training jobs, each can be performed in parallel.
The plotted processing rate is calculated by dividing the total size of the consumed data (number of epochs $\times$ dataset size) by the end-to-end runtime (including copying of the trained model to CPU's memory).
We observe that the FPGA scales until 14, since there are 14 engines, reaching 156~GB/s processing rate at the peak. When using 28 parallel threads, the XeonE5 can reach maximum 34~GB/s while POWER9 reaches 49~GB/s.

\begin{figure}[t]
\begin{tikzpicture}
\begin{axis}[
    xlabel=Time(s),
	ylabel=Logistic Loss,
	ymode=log,
	log ticks with fixed point,
    width=\columnwidth,
    height=4cm,
    xmin=0, xmax=0.75,
    ymin=0.00005, ymax=1,
    axis x line*=bottom,
    axis y line*=left,
    ymajorgrids,
    ytick={0.0001,0.001,0.01,0.1,1},
    y tick label style = {font=\scriptsize},
    xtick={0.1,0.2,0.3,0.4,0.5,0.6,0.7,0.8},
    x tick label style = {font=\scriptsize, rotate=0},
    every axis plot/.append style={ultra thick},
    legend columns=5,
    legend style={name=leg,at={(0.2,1.05)},anchor=south west,font=\scriptsize,draw=none}
]
\addplot[color=r4,mark=+] table [x index=1, y index=0] {plotdata/sgd_minibatch.dat};
\addplot[color=b4,mark=+] table [x index=2, y index=0] {plotdata/sgd_minibatch.dat};
\addplot[color=g4,mark=+] table [x index=3, y index=0] {plotdata/sgd_minibatch.dat};
\addplot[color=c4,mark=+] table [x index=4, y index=0] {plotdata/sgd_minibatch.dat};
\addplot[color=p4,mark=+] table [x index=5, y index=0] {plotdata/sgd_minibatch.dat};
\addlegendentry{1}
\addlegendentry{2}
\addlegendentry{4}
\addlegendentry{8}
\addlegendentry{16}
\end{axis}
\node[fill=white, draw=none, align=left, font=\scriptsize, below=0mm, anchor=east] (1) at (leg.west) {Minibatch Size:};
\end{tikzpicture}
\caption{Logistic loss over time depending on the minibatch on the FPGA with 1 engine. Dataset is \textit{IM}.}
\label{fig:sgd_minibatch}
\end{figure}
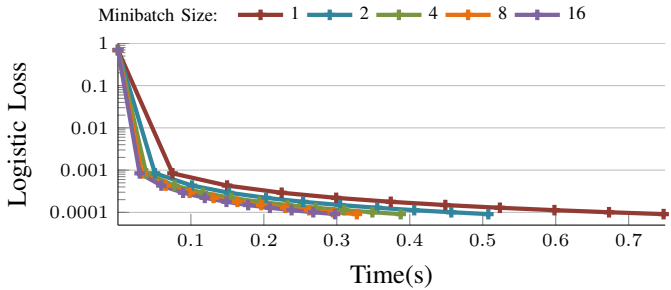

As in the microbenchmarks (Section~\ref{sec:background}), how the data is accessed is incredibly important for HBM performance: If there is only one copy of the dataset, it creates a bottleneck when accessing the HBM from 14 engines in parallel, diminishing the HBM advantage (Figure~\ref{fig:sgd_rate}): \textit{FPGA-nonreplicated} rate stays at 12.8~GB/s and does not scale with the number of engines used.
To utilize the HBM, we have to replicate the dataset such that each engine can access its own physical memory channel without utilizing the crossbar.
The replication will not be possible if the dataset is larger than 512~MiB (the size of 2 physical memory channels on the HBM, exposed as an AXI3 port by the HBM-shim in Section~\ref{sec:arch}).
However, in that case a blockwise scan approach can be followed~\cite{jaggi2014communication}:
A block of the dataset that fits 512~MiB is read for multiple epochs and then gets exchanged for the next block, and so on. This reduces the IO complexity between the CPU and the FPGA while replication leads to high bandwidth from HBM.

\noindent
\textbf{Evaluation: Effect of dimensionality.} In Figure~\ref{fig:sgd_datasets} we show how the processing rates across platforms and datasets differ. Lower dimensional datasets (\textit{AEA}) lead to lower processing rate per engine because the SGD pipeline (Figure~\ref{fig:sgd}) cannot be kept at 100\% utilization. The reason is that we keep the read-after-write (RAW) dependency (lines 4 and 7 in Algorithm~\ref{algo:sgd}) in SGD, causing bubbles during the processing if the dataset is low dimensional. In certain cases, the RAW dependency can be ignored in FPGA-based implementations~\cite{kara2017fpga}, since SGD tends to be tolerant against constant stale updates.
However, the tolerance is not guaranteed especially if the samples are ordered in a certain fashion, so we choose to respect the RAW dependency to get high quality convergence.

\noindent
\textbf{Evaluation: Minibatch size.} Figure~\ref{fig:sgd_minibatch} shows the convergence over time for minibatch sizes used during SGD on the FPGA.
Larger minibatch sizes increase pipeline utilization, leading to higher processing rate and therefore faster convergence.
The goal here is to show that we can converge to the same loss, when increasing the minibatch size from 1 to 16, while achieving significant speedup.
Accordingly, we use 16 in all SGD experiments both on the CPU and the FPGA (Figure~\ref{fig:sgd}).
\section{Discussion and Related Work}
\label{sec:discussion}

\begin{table}[]
\centering
\caption{Consumption on XCVU37P-2E-FSVH2892}
\label{tab:resources}
\setlength\tabcolsep{1.5pt}
\begin{tabular}{c|c|c|c|c|c|c|c}
\textbf{Bitstream} & \textbf{\#engines} & \textbf{LUT} & \textbf{LUTRAM} & \textbf{FF} & \textbf{BRAM} & \textbf{URAM} & \textbf{DSP} \\
\hline
Selection & 14 & 17.99\% & 3.35\% & 17.97\% & 26.53\% & 23.33\% & 0\% \\
Join & 7 & 40.81\% & 35.88\% & 26.13\% & 58.48\% & 23.33\% & 0\%\\
SGD & 14 & 55.76\% & 5.02\% & 47.29\% & 55.95\% & 46.66\% & 38.78\%\\
\end{tabular}
\end{table}

\noindent
\textbf{Discussion: Timing.} Although HBM offers obvious advantages for data analytics and machine learning as we have shown, there are major design challenges when implementing accelerators on HBM-based FPGAs.
Since taking advantage of HBM requires a scale-out architecture, the designs are more likely to be constrained on the available resources (Table~\ref{tab:resources}).
This entails the usage of multiple SLRs.
However, on Xilinx FPGAs all HBM ports are located at SLR0.
So, the AXI interfaces from the engines that are located in other SLRs need to be routed across SLRs to reach the HBM.
This makes the timing closure challenging and requires careful floorplanning.
The way we solved this issue to a certain extent is to first constrain one compute engine to be placed in just one SLR and second add AXI-interconnect modules with internal buffering in SLRs in between the compute engine and the HBM.
For instance, for a compute engine placed in SLR2, we put two AXI-interconnect modules in SLR1 and SLR0 to ease routing.
A possible solution to this problem might be hardened overlays on top of the FPGA fabric~\cite{achronix} that route signals in wide busses rather than bitwise routing.
As FPGAs become more popular for high performance data processing, efficient data movement becomes critical.
While HBM is a step in the right direction as we show in this work, more advancements on the device level are needed to make the development easier.

\noindent
\textbf{Discussion: Data movement.}
Coherent interconnects such as OpenCAPI~\cite{stuecheli2018ibm} allow uniform memory management between CPUs and accelerators: Address spaces on the accelerator can be mapped by the OS and used natively.
Due to the ongoing development of accelerator endpoints to enable this, for this work we used dedicated datamovers that need to be initiated separately by the software.
In the presence of uniform memory management, the integration to the database can be easier:  
The main copy of data can be mapped to the accelerator memory, eliminating initial copies,
making acceleration more interesting for data movement sensitive use cases such as the join.


\noindent
\textbf{HBM on many-core processors} are becoming common thanks to 3D-stacking~\cite{weis2011design} and chiplets. Prominent examples include processors such as Intel Knights Landing (KNL)~\cite{sodani2016knights}, NVIDIA Titan V, and Google's TPU~\cite{jouppi2017datacenter}.
Recent work has focused on showing the usefulness of HBM on KNL for data processing workloads.
KNL being an x86 many-core architecture offers easy portability for existing codebases and allows rapid testing of HBM-related ideas.
Cheng et al.~\cite{cheng2019deploying} focus on optimizing NUMA placement of hash tables on KNL to increase the utilization of the HBM and provide simulation results for hash join.
Pohl et al.~\cite{pohl2019joins} focus on using the HBM on KNL for joins and find that the mode where HBM is directly addressed as opposed to the cache-mode results in the highest performance.
Miao et al.~\cite{miao2019streambox} show how the HBM on KNL can be used to improve stream processing focusing on grouping operations, finding that sort-based algorithms fit HBM better than hash-based ones that require random access.

\noindent
\textbf{Database acceleration with FPGAs} is a popular field, where recent surveys have been published~\cite{papaphilippou2018accelerating,fang2019memory}.
Inherent parallelism and capability of creating specialized processing pipelines make FPGAs a good target in cases such as sorting~\cite{mueller2012sorting}, joins~\cite{casper2014hardware}, hashing~\cite{kara2016fast}, regular expression matching~\cite{sidler2017accelerating}, high fanout data partitioning~\cite{kara2017fpga2}, compression~\cite{fowers2015scalable}, and grouping in the datapath~\cite{woods2014ibex}.
Accelerator integration is a challenge that has been loosely addressed by systems such as doppioDB~\cite{alonso2019doppiodb}.
When considering end-to-end integration of FPGA-based accelerators, HBM might not be advantageous at all, especially if the accelerator relies on hybrid processing~\cite{kara2017fpga2}, so in this work we considered cases that can be performed completely on the FPGA, taking advantage of the HBM.
Other workloads such as sorting and grouping might benefit from HBM just as well, following similar principles.

\noindent
\textbf{In-database machine learning} is nowadays commonly performed~\cite{db2ml,oracleml}.
Recent work~\cite{mahajan2018rdbms,wang2019accelerating,kara2018columnml} has focused on improving these operations with FPGAs,
with an additional benefit of an offload from the CPU, leaving the CPU better suited 
for processing database queries with low response times~\cite{kara2019doppiodb}.
Mahajan et al.~\cite{mahajan2018rdbms} propose a framework for FPGA accelerators to perform training with SGD and low-rank matrix factorization within PostgreSQL.
Wang et al.~\cite{wang2019accelerating} propose a fully specialized pipeline to take advantage of quantized datasets when training with SGD.
Kara et al.~\cite{kara2018columnml} focus on training
using column-stores
and perform on-the-fly decompression/decryption on the FPGA.
Thanks to the HBM, we are able to surpass their performance substantially at 156 GB/s, compared to 3.76 GB/s~\cite{mahajan2018rdbms} and 15 GB/s~\cite{wang2019accelerating} under same configurations (SGD with logistic regression and full precision input data).
When implemented using the same principles, these works would likely benefit from HBM just as well, where resource consumption will be the determining factor to reach the target scale-out parallelism.

\section{Conclusion}

We show an end-to-end system with HBM-attached FPGA accelerators integrated into a main-memory database.
Thanks to the HBM, even memory-bound algorithms such as range selection benefit from being implemented on the FPGA.
Data intensive algorithms such as joins benefit under certain conditions, while offloading machine learning workloads such as SGD that access data iteratively provides consistent improvement compared to CPU-based solutions.
For all workloads, we have shown the importance of data partitioning and address space usage in the HBM to utilize it to its full potential.

\bibliographystyle{IEEEtran}
\bibliography{IEEEabrv,main}

\end{document}